\documentclass[Physsubmission, Phys]{SciPost}
\hypersetup{
    colorlinks,
    linkcolor={red!50!black},
    citecolor={blue!50!black},
    urlcolor={blue!80!black}
}

\usepackage[bitstream-charter]{mathdesign}
\usepackage{lineno}
\usepackage{array}
\usepackage{multirow}
\DeclareSymbolFont{usualmathcal}{OMS}{cmsy}{m}{n}
\DeclareSymbolFontAlphabet{\mathcal}{usualmathcal}

\begin{document}

\begin{center}{\Large \textbf{
%Article Title, as descriptive as possible, ideally fitting in two lines (approximately 150 characters) or less\\
PDF4LHC21: Update on the benchmarking of the CT, MSHT and NNPDF global PDF fits
}}\end{center}

% TODO: write the author list here. Use initials + surname format.
% Separate subsequent authors by a comma, omit comma at the end of the list.
% Mark the corresponding author with a superscript *.
\begin{center}
Thomas Cridge\textsuperscript{1} on behalf of the PDF4LHC21 combination group.
%Dee E. Faa\textsuperscript{1},
%Aah B. Cee\textsuperscript{2} and
%Gee K. See\textsuperscript{3$\star$}
\end{center}

\begin{center}
{\bf 1} Department of Physics and Astronomy, University College London, London, WC1E 6BT, UK
%\\
%{\bf 2} Affiliation2
%\\
%{\bf 3} Affiliation2
%\\
t.cridge@ucl.ac.uk
\end{center}

\begin{center}
\today
\end{center}

% For convenience during refereeing (optional),
% you can turn on line numbers by uncommenting the next line:
%\linenumbers

\definecolor{palegray}{gray}{0.95}
\begin{center}
\colorbox{palegray}{
  \begin{tabular}{rr}
  \begin{minipage}{0.1\textwidth}
    \includegraphics[width=22mm]{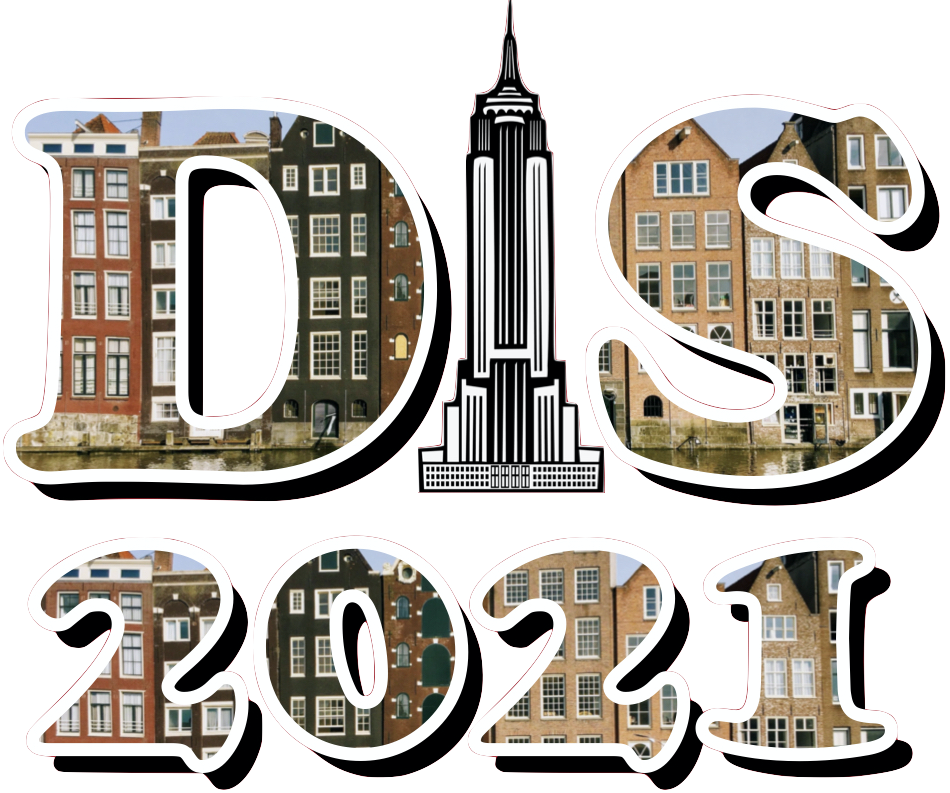}
  \end{minipage}
  &
  \begin{minipage}{0.75\textwidth}
    \begin{center}
    {\it Proceedings for the XXVIII International Workshop\\ on Deep-Inelastic Scattering and
Related Subjects,}\\
    {\it Stony Brook University, New York, USA, 12-16 April 2021} \\
    \doi{10.21468/SciPostPhysProc.?}\\
    \end{center}
  \end{minipage}
\end{tabular}
}
\end{center}

\section*{Abstract}
{\bf
% TODO: write your abstract here.
%The abstract is in boldface, and should fit in 8 lines.
%It should be written in a clear and accessible style, emphasizing the context, the problem(s) studied, the methods used, the results obtained, the conclusions reached, and the outlook. You can add a table contents, recommended if your paper is more than 6 pages long.

There have been recent updates to the three global PDF fits (CT, MSHT and NNPDF), all adding large amounts of data from the LHC, and this has resulted in significant changes to the global PDFs. Given the impact that the new PDFs will have on physics comparisons at the LHC, it is crucial to perform a benchmarking among the PDFs, similar in spirit to that which was carried out for PDF4LHC15, widely used for LHC physics. In this article we detail a benchmarking comparison of three global PDF sets - CT18, MSHT20 and NNPDF3.1 - and their similarities and differences that have been observed. The end result of this study will be a new PDF4LHC21 ensemble of combined PDFs suitable for a wide range of LHC applications.

%The end result of this exercise will be a PDF4LHC21 set of PDFs, formed from the combination of the three global PDF sets: CT18, MSHT20, NNPDF3.1. In this article we detail the benchmarking that has been performed, the similarities and differences observed, and determine the origin of any such differences. This has an impact on both the understanding of the global fit PDFs and on the future combination PDF4LHC21 PDFs for use at the LHC.
}

\vspace{10pt}
\noindent\rule{\textwidth}{1pt}
\tableofcontents\thispagestyle{fancy}
\noindent\rule{\textwidth}{1pt}
\vspace{10pt}

\section{Introduction}
\label{sec:intro}

The knowledge and determination of the parton distribution functions (PDFs) of the proton is a key requirement for the physics program at the Large Hadron Collider (LHC) and future colliders. Both the central values of the PDFs and their uncertainties play an important role in many analyses. As a result, the accurate and precise evaluation of PDFs from available experimental data is of paramount importance. Over the past few years there have been many developments in this area, with the inclusion of processes at NNLO in PDF fits, a great variety of new data from the LHC now included and continuing developments of the PDF fitting methodologies. Consequently, the PDFs are now known more accurately and precisely than ever before, with the global fitting groups having recently provided major updates and new PDF sets including these advances \cite{MSHT20,CT18,NNPDF3.1}. 

However, with these advances, differences have begun to emerge in both the central values and the uncertainties of the PDFs extracted by different groups. As the experimental and PDF uncertainties continue to reduce, such differences may become a key component of the uncertainties unless they can be understood, particularly in the context of a new PDF4LHC combination set. Moreover, with statistical precision increasing for many measurements, challenges have arisen in fitting some of the datasets available, and differences in the fitting methodology (e.g. in the treatment of correlated systematics) may have effects on the PDFs. Given therefore that it has been several years since the last major benchmarking effort of global PDF fits in PDF4LHC15 \cite{PDF4LHC15} (itself the last in a line of several efforts including 
\cite{Alekhin:2011sk,Ball:2012wy,SM:2010nsa,Rojo:2015acz}), the time is ripe for a study of these differences. We present such a benchmarking exercise in these proceedings. The need is made all the more pressing by the upcoming LHC Run III, which will provide yet further information constraining the PDFs and also utilise the PDFs as a key input in physics analyses.

% \cite{MMHTjets,Bailey:2019yze,Harland-Lang:2017ytb,Thorne:2019mpt,Boughezal:2017nla,AbdulKhalek:2020jut,Amat:2019upj,Hou:2019gfw,ATLAS8jets,ATL-PHYS-PUB-2018-017,Amoroso:2020lgh}

In these proceedings we describe the progress made so far in benchmarking three of the global PDF fits: CT, MSHT and NNPDF, with our aims being first to identify differences where they arise and then to understand the origin of these differences through their detailed examination. Ultimately these comparisons will lead to an improved understanding of the differences in the PDFs and enable the production of a new PDF4LHC combination ensemble, to be reported in a future publication. This will be a combination of the most recent versions of the global fits of the three groups: CT18, MSHT20 and NNPDF3.1\footnote{Specifically, it will be an updated and modified version of NNPDF3.1, similar to that used in the NNPDF study of the strangeness \cite{Faura:2020oom}.} in the style of PDF4LHC15, reflecting the current knowledge of the PDF central values, their uncertainties and spread for use in the LHC at Run III and beyond, as well as wherever else is appropriate.

\section{Approach to Benchmarking}
\label{sec:approach}

In order to determine any differences in the PDFs, we seek to begin with a baseline of PDF fits from the three groups where as many differences in input and approach are removed as possible, this allows the benchmarking to focus upon salient differences. We therefore choose both a common input set of datasets to include and common theory settings. This produces the so-called \textit{``reduced fits''}, offering ease of comparison at the expense of the usual robustness offered by global fits. 

We begin with the choice of input data, to set up the reduced fits we keep only the largest subset of data fit by all three groups in an (almost) identical manner. Furthermore, we apply the most conservative cuts made by any group, i.e. $Q^2 > 4 {\rm GeV}^2$ and $W^2 > 15 {\rm GeV}^2$. Given numerous subtle differences between the groups, the final list of common data for the reduced fit is rather small and shown in Table~\ref{tab:datasets}. This reduced fit set of datasets also satisfies the competing requirement of being sufficiently large and varied so as to provide some constraints on all the PDF combinations and their uncertainties. The constraints placed by this reduced fit dataset will necessarily be significantly more limited than in the usual full global fits, but we trade the robustness of the global fits for the ease of comparison of the reduced fits for the purposes of benchmarking.
%
%As a result, we still fit data from older fixed target DIS experiments, such as BCDMS (cite) and NMC, the crucial full HERA combined dataset is also included, whilst the NuTeV dimuon data is included to constrain the strangeness. Finally, newer LHC data on Drell-Yan, including the important high precision ATLAS 7~TeV $W,Z$ data, is included, whilst the CMS 8~TeV inclusive jet data constrains the gluon at high $x$.

\begin{table}[t]
\begin{center}
\fontsize{9}{11}\selectfont 
\renewcommand\arraystretch{1.0}
%\vspace{-1.0cm}
\begin{tabular}{|c|c|c|c|}
\hline Dataset & Reference & Dataset & Reference \\ \hline
BCDMS proton, deuteron DIS & \cite{BCDMS} & LHCb 8TeV $Z \rightarrow ee$ & \cite{LHCb-Zee}\\ 
NMC deuteron to proton ratio DIS & \cite{NMCn/p} & ATLAS 7~TeV high precision $W,Z$ (2016) & \cite{ATLASWZ7f} \\  
NuTeV $\nu N$ dimuon & \cite{Dimuon} & D0 $Z$ rapidity & \cite{D0Zrap} \\ 
HERA I+II inclusive DIS & \cite{H1+ZEUS}\cite{HERAcomb} & CMS 7~TeV electron asymmetry & \cite{CMS-easym}\\ 
E866 Drell-Yan ratio $pd/pp$ DIS & \cite{E866DYrat} & ATLAS 7~TeV $W,Z$ rapidity (2011) & \cite{ATLASWZ} \\ 
LHCb 7, 8~TeV $W,Z$ rapidity & \cite{LHCb-WZ} & CMS 8~TeV inclusive jet & \cite{CMS8jets} \\ 
\hline
    \end{tabular}
\end{center}
\caption{\sf Datasets included in the initial baseline reduced fits.}
\label{tab:datasets}   
\end{table}

With the differences in the input data now removed, we also choose common theory settings wherever possible\footnote{However each group still uses their own version of the theoretical calculation at the appropriate order in QCD.} including: the same heavy quark masses and strong coupling ($m_c^{\rm pole} = 1.4 {\rm GeV}$, $m_b^{\rm pole} = 4.75 {\rm GeV}$ and $\alpha_S(M_Z^2) = 0.118$); setting no strangeness asymmetry at the input scale, i.e. $(s-\bar{s})(Q_0) = 0$; perturbative charm; positive definite quark distributions; no deuteron or nuclear corrections; and for the dimuon data we take a fixed branching ratio for charm hadrons to muons $B(D\rightarrow \mu)$ (this will be elucidated further later in this article in the section describing the dedicated study of the reduced fit strangeness) and NNLO corrections.

%Common theory settings:
%\begin{itemize}
%\item Same heavy quark masses ($m_c = 1.4 {\rm GeV}$, $m_b = 4.75 {\rm GeV}$) and strong coupling $\alpha_S(M_Z^2) = 0.118$.
%\item No strangeness asymmetry at input scale, i.e. $(s-\bar{s})(Q_0) = 0$.
%\item Perturbative charm.
%\item Positive definite quark distributions.
%\item No deuteron or nuclear corrections.
%\item Fixed branching ratio for charm hadrons to muons $B(D\rightarrow \mu)$.
%\item NNLO corrections for dimuon data. 
%\end{itemize}

%The first of these is an obvious source of potential valid differences between fits. The lack of strangeness asymmetry and requirement of perturbative charm remove further differences in approach taken by individual global fitting groups, whilst the requirement of positive definite quark distributions can be important when dealing with such reduced datasets. No deuteron or nuclear corrections are applied as whilst all 3 groups can now apply these, they do so in different ways (cite), nonetheless differences are not expected to be large from this source. Finally, the last two requirements of a fixed branching ratio for charm hadrons to muons and NNLO corrections for the dimuon data apply to the NuTeV, and relate to specific differences discussed in Section~\ref{sec:strangeness}.

At this point it is worth explicitly noting that neither the reduced fit dataset, nor common theory settings for the reduced fit, correspond to the data or theory settings used by any group, rather they are a compromise to the least common denominator in each case. Indeed, some of the choices made are known to be sub-optimal but we adopt them for the purposes of the benchmarking exercise only. Even when these differences in input and theory are removed, methodological differences remain, including the choice of the General-Mass Variable-Flavour-Number scheme and also the treatment of the PDF uncertainties. These will not be described here but the former has been compared previously in \cite{SM:2010nsa,Guzzi:2011ew,Andersen:2014efa} whilst the latter will be outlined in our future full publication of the benchmarking exercise.

\section{Comparison of Reduced and Global Fits}
\label{sec:reducedvsglobal}

\begin{figure}[h]
\centering
\includegraphics[width=0.49\textwidth,trim= 0.5cm 0.5cm 0.3cm 0.4cm,clip]{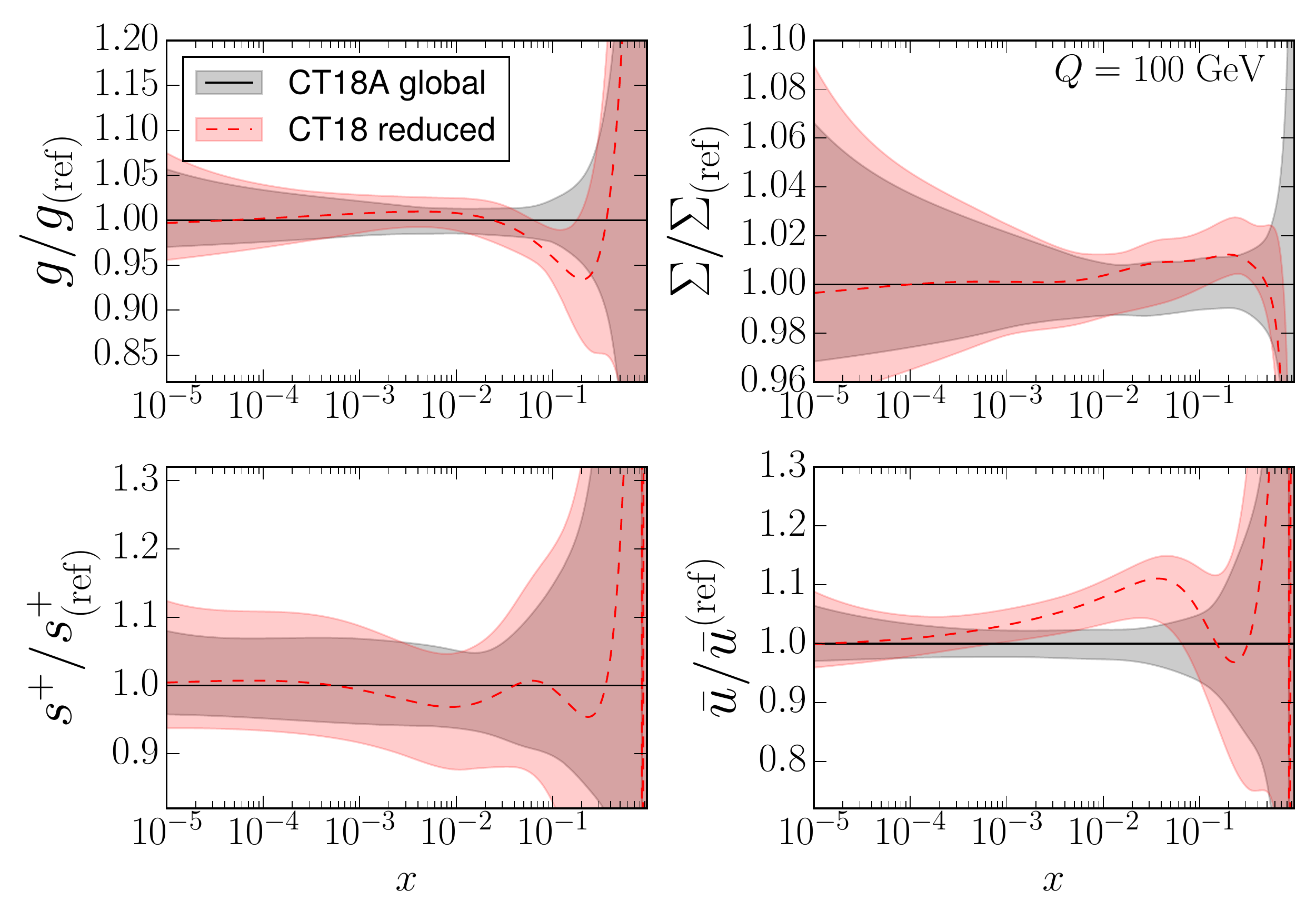}
\includegraphics[width=0.49\textwidth,trim= 0.5cm 0.5cm 0.3cm 0.4cm,clip]{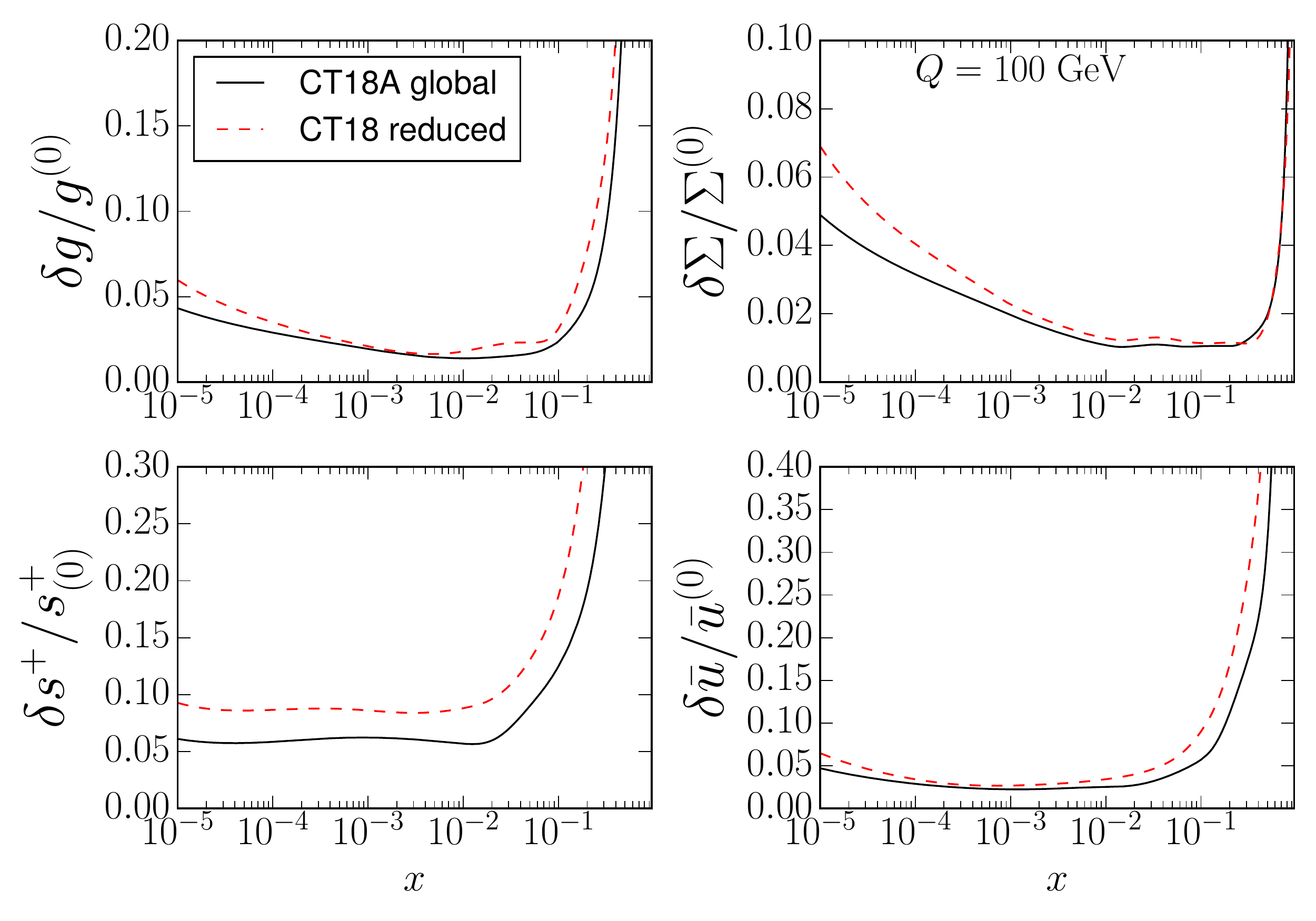}
\caption{Comparison of the CT reduced fit relative to the CT18A global fit, the ratio of the central values is shown in the left plots and the uncertainties are compared in the right plots. The gluon, singlet ($\Sigma$), total strangeness ($s^+=s+\bar{s}$) and anti-up ($\bar{u}$) are compared. The baseline for the ratio is the CT18A global fit.}
\label{fig:CTreducedvsglobal_ratio}
\end{figure}

At this stage, before performing the benchmarking of the reduced fits, it is useful to compare the reduced fits produced by each of the three groups with their default global fits. We show this comparison for CT only for brevity in Figure~\ref{fig:CTreducedvsglobal_ratio}. Overall good compatibility is observed between the central values in the ratio of the CT18 reduced fit to the CT18A global fit\footnote{CT18A is compared against as this includes the ATLAS 7~TeV $W,Z$ data, which is included in the reduced fit.}, with changes in the high-$x$ gluon shape resulting from the reduced amount of jet and other data relevant in this region in the reduced fit. The singlet and strangeness are both compatible within the uncertainties of the global fit, whilst there is an increase in the anti-up PDF at intermediate $x$, which signals a change in the flavour decomposition in the reduced fit. Such changes are not unexpected given the significantly altered, and specifically reduced, dataset of the reduced fit. In the right-hand plot of Figure~\ref{fig:CTreducedvsglobal_ratio} the size of the uncertainty bands are compared, clearly there is some increase in the nominal PDF uncertainties of the reduced fit, particularly at low $x$. Generally, an increase in the PDF uncertainties (as evaluated in an identical manner to the global fit) is to be expected given the significantly reduced amount of data constraining the PDFs. Similar differences in central values and uncertainties are seen in comparing the MSHT and NNPDF reduced fits with their respective global fits, albeit MSHT sees a reduced strangeness relative to their global fit due to the exclusion of the ATLAS 8~TeV $W,Z$ data \cite{ATLAS8Z3D,ATLASW8} from the reduced fit. All reduced fits are compared in the next section where the benchmarking of the reduced fits is performed.

%, this data is present in the MSHT20 global fit and was shown to increase the strangeness \cite{MSHT20}

\section{Benchmarking of Reduced Fits}
\label{sec:benchmarking}

%There are three main methods that have been used and we demonstrate two of these in this article. 

\begin{figure}[h]
\centering
\includegraphics[width=0.49\textwidth,trim= 0.5cm 0.5cm 0.3cm 0.4cm,clip]{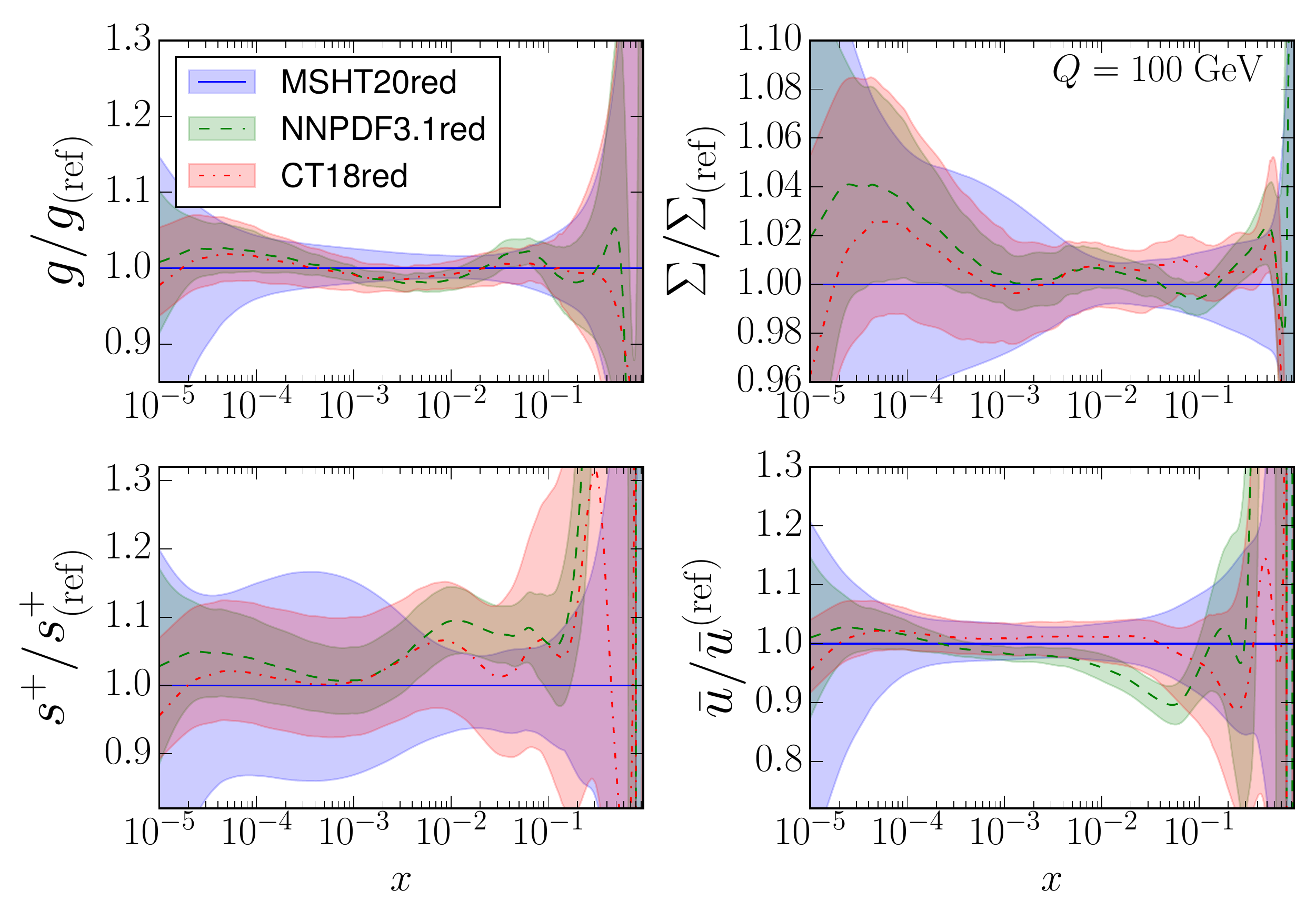}
\includegraphics[width=0.49\textwidth,trim= 0.5cm 0.5cm 0.3cm 0.4cm,clip]{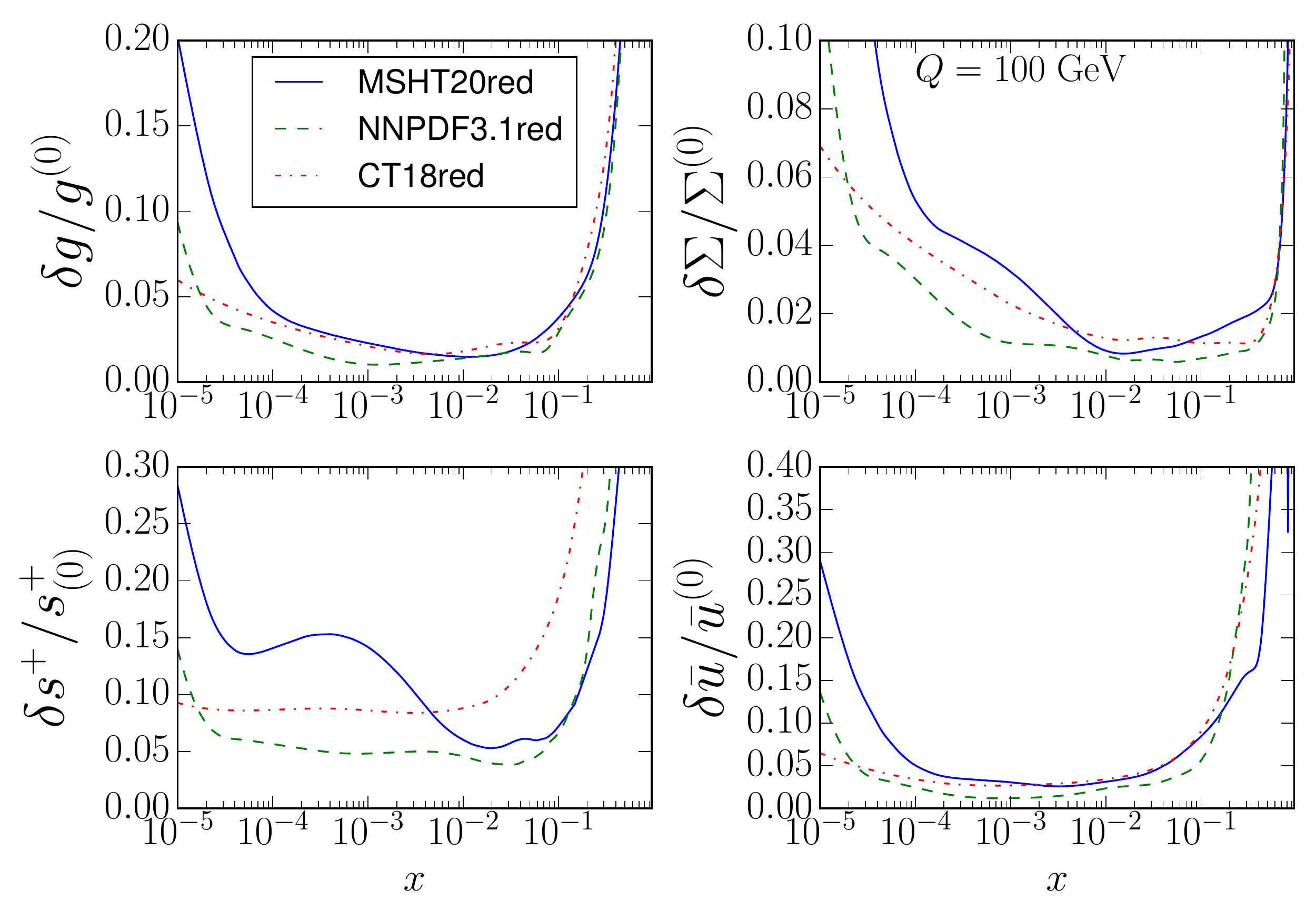}
\caption{Comparison of the reduced fits of all 3 groups (CT, NNPDF, MSHT), ratio of central values (left) and uncertainties (right), with the baseline for the ratio taken as the MSHT20 reduced fit.}
\label{fig:reducedvsglobal_ratio}
\end{figure}

Now that we have a gauge of the differences between the reduced fits and the global fits, we begin the benchmarking of the reduced fits. Firstly we compare the reduced fit PDFs themselves directly against one another at the level of both central values and uncertainties, as presented in Figure~\ref{fig:reducedvsglobal_ratio}. There is good general agreement between the three reduced fits within uncertainties for the majority of the PDFs over most of the $x$ range. Starting with the gluon, all three groups agree (almost) within uncertainties over the entirety of the $x$ range, this confirms that changes in the high-$x$ gluon shape relative to the global fits are driven by the datasets included. The singlet is also in very good agreement for all $x$. The strangeness is also largely consistent, albeit the NNPDF3.1 reduced fit is notably high around $10^{-2} \lesssim x \lesssim 10^{-1}$ and outside of the MSHT20 reduced fit uncertainty bands here (although within the CT18 reduced fit uncertainty bands). This difference in the strangeness will be examined later in the dedicated studies on the strangeness of the reduced fits. The NNPDF $\bar{u}$ is lower than both MSHT and CT in the same region, signalling a difference in the high-$x$ flavour decomposition. The uncertainties of the three reduced fits are also similar in size in data regions, with the MSHT reduced fit having larger uncertainties outside of these regions, i.e. where constraints are lacking in the reduced fit - particularly at low $x$. A parallel study into differences in the uncertainty bands, both the methodologies of their evaluation and differences in the reduced fit uncertainties, is ongoing and will be reported in the future.

%\begin{figure}[h]
%\centering
%\includegraphics[width=0.61\textwidth,trim= 0.5cm 0.5cm 0.3cm 0.4cm,clip]{figures/pdfplot-err-allcomp-June21-q100Gev_large.pdf}
%\caption{Comparison of the uncertainties of the reduced fits of all 3 groups (CT, NNPDF, MSHT).}
%\label{fig:reducedvsglobal_errs}
%\end{figure}

%; secondly we compare the reduced fits at the level of the $\chi^2$ at the dataset by dataset level. The latter method is especially useful in identifying specific datasets causing observed differences, where such differences were seen, data and theory predictions themselves were directly compared to focus on the origin of the differences, although this will not be presented here. All of these comparisons can be further augmented by first using fixed PDFs, before fitting, to remove this source of difference. More detailed benchmarking comparisons will be presented in a future publication.

In order to further uncover any differences in the reduced fits we may determine the dataset by dataset fit qualities, as given by the $\chi^2/N_{\rm pts}$. Before calculating the fit qualities for the reduced fits, it is useful to compare them before fitting, i.e. with fixed (PDF4LHC15 NNLO) PDFs. This allows any differences of theory or data between the groups to be identified before the complication of fitting is added. Overall, for most of the reduced fit datasets there is good agreement observed in the size of the $\chi^2/N_\mathit{pts}$ before fitting, reflecting agreement in the data/theory comparison for the three groups. Therefore we can progress to comparing the reduced fits' fit qualities after fitting, this is shown in Table~\ref{tab:reducedfit_chisqs_fit}. Very good overall agreement is observed between the three reduced fits with overall fit qualities in terms of $\chi^2/N_\mathit{pts}$ of 1.14, 1.15, 1.20 for the CT, MSHT and NNPDF reduced fits respectively. Moreover, this global agreement follows on into the dataset by dataset comparisons with the majority showing good agreement. There are nonetheless some differences, the NuTeV dimuon data shows a noticeable difference in fit quality between CT and MSHT which each observe 0.8 per point, and NNPDF which observes 1.2 per point. This is consistent with the increased strangeness noted earlier in the NNPDF reduced fit in the region $10^{-2} \lesssim x \lesssim 10^{-1}$ as the dimuon data favours lower strangeness here. However this occurs without improvement of the ATLAS 7~TeV $W,Z$ data which pulls the strangeness upwards in this $x$ region. The origin of this difference in the strangeness will be investigated in the first section of the dedicated studies later in this article. An additional difference observed is in the NNPDF fit quality to smaller datasets, such as the E866 Drell-Yan ratio data or the CMS 7~TeV electron asymmetry. For the former there is a known difference between CT and MSHT due to the parameterisation of the $\bar{d},\bar{u}$ ratio or difference in this high-$x$ region (\cite{MSHT20},\cite{CT18}). For the latter, CT and MSHT both obtain 1.5 per point whereas NNPDF obtains half of this at 0.76, again this will be analysed briefly later in the context of the effect of different weights in the PDF fits in the dedicated study of the high-$x$ gluon.

\begin{table}[t]
\fontsize{9}{11}\selectfont 
  \renewcommand\arraystretch{1.0} 
\centering  \begin{tabular}{llrrrrr}
%  \renewcommand\arraystretch{0.9}
  %\cline{1-1}
  %\multicolumn{1}{|l|}{}  
  ID       & Expt.                               &  $N_\mathit{pts}$ & $\chi^2/N_\mathit{pts}$ (CT) & $\chi^2/N_\mathit{pts}$ (MSHT) & $\chi^2/N_\mathit{pts}$ (NNPDF)     \\
  \hline             
  101      & BCDMS $F_2^p$                       &  329/163$^{\dagger\dagger}$/325$^\dagger$   &    1.06   &   1.00   &   1.21  \\
  102      & BCDMS $F_2^d$                       &  246/151$^{\dagger\dagger}$/244$^\dagger$   &    1.06   &   0.88   &   1.10  \\
  104      & NMC $F_2^d/F_2^p$                   &  118/117$^\dagger$                          &    0.93   &   0.93   &   0.90  \\
  124+125  & NuTeV $\nu \mu\mu+\bar{\nu} \mu\mu$ &  38+33                                      &    0.79   &   0.83   &   1.22  \\
  160      & HERAI+II                            &   1120                                      &    1.23   &   1.20   &   1.22  \\
  203      & E866 $\sigma_{pd}/(2\sigma_{pp})$   &     15                                      &    1.24   &   0.80   &   0.43  \\
  245+250  & LHCb 7TeV \& 8TeV  $W$,$Z$           &  29+30                                      &    1.15   &   1.17   &   1.44  \\
  246      & LHCb 8TeV $Z \rightarrow ee$        &     17                                      &    1.35   &   1.43   &   1.57  \\
  248      & ATLAS 7TeV $W$,$Z$(2016)            &     34                                      &    1.96   &   1.79   &   2.33  \\
  260      & D0 Z rapidity                       &     28                                      &    0.56   &   0.58   &   0.62  \\
  267      & CMS 7TeV electron $A_{ch}$           &     11                                      &    1.47   &   1.52   &   0.76  \\
  269      & ATLAS 7TeV $W$,$Z$(2011)            &     30                                      &    1.03   &   0.93   &   1.01  \\
  545      & CMS 8TeV incl. jet                  &  185/174$^{\dagger\dagger}$                 &    1.03   &   1.39   &   1.30  \\
  \hline                                                                                    
  \hline                                                                                    
  Total    & $N_\mathit{pts}$                     &    ---    &   2263       &    1991       &   2256   \\     
  \hline                                                                                    
  Total    & $\chi^2/N_\mathit{pts}$              &    ---    &   1.14       &    1.15       &   1.20   \\        
  \end{tabular}
          \caption[]{PDF4LHC21 reduced fit dataset $\chi^2/N_\mathit{pts}$ after fitting, $^{\dagger\dagger}$MSHT\ \ $^{\dagger}$NNPDF.}
\label{tab:reducedfit_chisqs_fit}                                
\end{table}

\section{Dedicated Studies}
\label{sec:specifics}

Overall we have now seen that the reduced fits of the three groups are in good agreement in both the PDFs and fit qualities of most datasets. Nonetheless we highlighted some differences which we will now scrutinise alongside other areas of interest to the global PDF fits.

\subsection{Strangeness and NuTeV}
\label{sec:strangeness}

\begin{figure}[h]
\centering
\includegraphics[width=0.54\textwidth,trim= 0.5cm 0.5cm 0.3cm 0.0cm,clip]{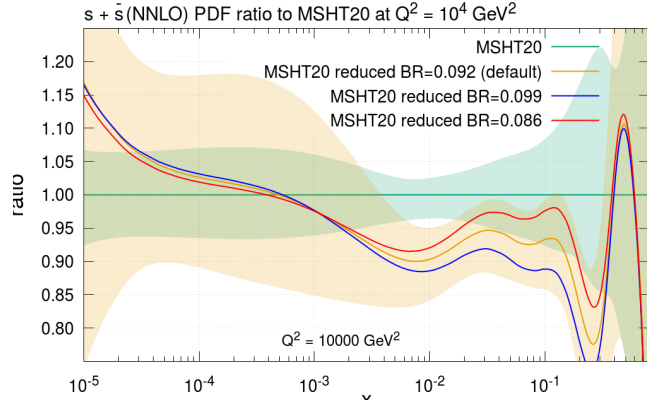}
\caption{Effect of the charm hadrons to muon branching ratio (BR) $B (D \rightarrow \mu)$ on the total strangeness of the MSHT reduced fit, relative to the MSHT20 global fit baseline.}
\label{fig:MSHTstrangeness_dimuonBR}
\end{figure}

One of the main differences emerging both in the PDFs and the dataset $\chi^2$ for the reduced fits was in the strangeness, $s+\bar{s}$. The main datasets driving this in the reduced fits and the global fits are the NuTeV dimuon data and the ATLAS 7~TeV $W,Z$ data (\cite{MSHT20,CT18,NNPDF3.1}, \cite{Aaboud:2016btc,Thorne:2017aoa,Faura:2020oom}), with the former preferring reduced strangeness at intermediate $x$ and the latter favouring enhanced strangeness in this region. The NuTeV dataset has several complications, requiring treatment of the non-isoscalar nature of the iron target, acceptance corrections and knowledge of the charm hadrons to muons branching ratio  \cite{Mason:2006qa}, $\rm {BR}(c \rightarrow \mu)$. The first two of these are treated consistently between the three fits, the last is needed as this process involves the scattering of a muon neutrino off an iron nucleon, this directly produces a muon but also results in the production of a charm hadron which decays to produce a further muon (hence dimuon). By default the three groups use different values for this branching ratio, NNPDF take a PDG value of $0.087\pm 0.005$ \cite{PDG2020}, MSHT take a value of $0.092\pm 0.01$ taken from direct measurements \cite{Harland-Lang:2014zoa}, \cite{Bolton:1997pq}\footnote{Taking the value from direct measurements avoids a potential circularity in its determination and use.} and further allow the central value to vary with a penalty within this range, and CT take the value of 0.099 used by NuTeV itself \cite{Mason:2007zz} with a normalisation uncertainty. However, the total strangeness obtained in the data region $10^{-2} \lesssim x \lesssim 0.4$ is anti-correlated with the branching ratio as it maps from the total strangeness to the NuTeV predictions in the fit. Therefore a smaller branching ratio allows one to fit the same NuTeV data whilst retaining larger strangeness and vice versa. The effect of taking the different branching ratios of the three groups within the MSHT reduced fit is shown in Figure~\ref{fig:MSHTstrangeness_dimuonBR}, where an increase/decrease in the total strangeness relative to the default value of approximately 5\% is observed as the branching ratio is altered. Upon removal of this difference, the NNPDF strangeness reduces whilst the CT strangeness increases as expected, bringing the reduced fits into closer agreement. Given the lack of constraints in a reduced fit, this improved strangeness agreement also enables a reduction in differences in the flavour decomposition.

\subsection{High-$x$ gluon, top and inclusive jet data}
\label{sec:highxgluon}

Another area of interest for the reduced fits and global fits is the high-$x$ gluon. We saw in the comparison of the reduced fit gluon in Figure~\ref{fig:reducedvsglobal_ratio} that the gluon was consistent between the three reduced fits. However, this contrasts the situation for the global fits where notable differences are present depending on the datasets included, their treatment and relative weights. This is further complicated by various issues in fitting many of these datasets including difficulties fitting all bins, possible tensions within and between dataset types and issues of correlated systematics. Therefore this is an obvious region that would benefit from benchmarking. There are three main dataset types which play a role in this region, all largely coming from the LHC in the past few years, these are jet data, top data and $Z$ $p_T$ data.

%\citeMMHTjets,Bailey:2019yze,Harland-Lang:2017ytb,Thorne:2019mpt,Boughezal:2017nla,AbdulKhalek:2020jut,Amat:2019upj,Hou:2019gfw,ATLAS8jets,ATL-PHYS-PUB-2018-017,Amoroso:2020lgh
% We wish to understand any issues and differences better by examining this in the context of the reduced fits.

%and the need for decorrelations 

The ATLAS 8~TeV multi-differential $t\bar{t}$ lepton + jets data \cite{ATLASsdtop} is one such dataset. This dataset has been studied by all three groups  \cite{MSHT20}, \cite{CT18}, \cite{Bailey:2019yze,Amat:2019upj,Hou:2019gfw,Kadir:2020yml,Czakon:2016olj,Harland-Lang:2017ytb,Thorne:2019mpt,AbdulKhalek:2020jut} and comes differential in four variables: $m_{tt}, y_{t}, y_{tt}$ and $p_t^T$, with statistical and systematic correlations provided to enable all four distributions to be fitted together. However several groups, including CT and MSHT, observed difficulties in fitting all four distributions simultaneously, achieving $\chi^2/N_\mathit{pts} \gtrsim 5$. Moreover both groups and ATLAS found issues fitting the rapidity distributions even individually \cite{ATL-PHYS-PUB-2018-017,Amoroso:2020lgh}. On the other hand, NNPDF3.0 were able to fit all four distributions individually\cite{Czakon:2016olj}\footnote{However, the statistical and systematic correlations were not available at the time of the NNPDF3.0 analysis.}. In order to analyse the differences arising here we therefore add this to our reduced fit baseline, this is shown in Table~\ref{tab:ttbar_chisqs}. First we check the theory predictions and data for fixed (PDF4LHC15) PDFs, this verifies that the data and theory are in agreement between the three groups. Moreover, at this stage all three groups observe the same pattern in the individual distributions' $\chi^2$, with all groups unable to obtain a good fit to the rapidity distributions.

After fitting however, differences emerge. If all four distributions are added but without any correlations (statistical or systematic) between the distributions - the ``uncorrelated case''\footnote{This ``uncorrelated'' case is not how this data should be added to the fit, however it enables the fit qualities of the individual distributions to be observed which offers useful information for understanding the differences.} - then whilst CT and MSHT retain the same behaviour, NNPDF are now able to fit the rapidity distributions well, albeit at the expense of a worsening of the fit quality to the $p_t^T$. On the other hand, if all statistical and systematic correlations between the four distributions are retained - the ``correlated case'' - then both MSHT and NNPDF again observe similar behaviour with neither reduced fit able to fit the 4 distributions together, as is seen in the global fits.

% Finally, in the last row of Table~\ref{tab:ttbar_chisqs} we show that by partially decorrelating the parton shower systematic between and within distributions in the same way as in the MSHT20 default global PDF fit (cite,cite) then a reasonable fit quality can be recovered.

\begin{table}[t]
\fontsize{8.2}{10.2}\selectfont
  \renewcommand\arraystretch{0.9} 
  \centering
  \begin{tabular}{|>{\centering\arraybackslash}m{5.4cm}|l|c|c|c|c|c|}
%  \renewcommand\arraystretch{0.9}
  %\cline{1-1}
  %\multicolumn{1}{|l|}{}  
   \hline
	Distribution/$N_\mathit{pts}$ & Group & $p_t^T/8$ & $y_t/5$ & $y_{tt}/5$ & $m_{tt}/7$ & Total/25 \\ \hline
\multirow{3}{5.4cm}{\centering PDF4LHC15 read in \\ default (CMS 8~TeV jets)} & MSHT & 3.0 & 10.6 & 17.6 & 4.3 & 35.5  \\ 
													  & CT & 3.1 & 10.1 & 15.3 & 4.2 & 32.7 \\ 
													  & NNPDF & 3.4 & 9.5 & 16.2 & 4.2 & 33.3 \\ \hline 
\multirow{3}{5.4cm}{\centering Uncorrelated case \\ default (CMS 8~TeV jets)} & MSHT & 3.8 & 8.4 & 12.5 & 6.4 & 31.2  \\ 
													  & CT & 3.4 & 12.9 & 17.3 & 6.1 & 39.7 \\ 
													  & NNPDF & 7.2 & 3.9 & 5.1 & 2.5 & 18.7 \\ \hline 
\multirow{2}{5.4cm}{\centering Correlated case \\ default (CMS 8~TeV jets)}    & MSHT & - & - & - & - & 130.6  \\ 
													  & NNPDF & - & - & - & - & 122.7 \\ \hline 
%MSHT default (partial decorrelation)                  & MSHT & - & - & - & - & 35.3 \\ \hline 
\centering Uncorrelated case double weight \\  (CMS 8~TeV jets) & MSHT & 4.2 & 5.8 & 7.4 & 6.5 & 23.9  \\  \hline 
\centering Uncorrelated case \\ (no LHC jets) & MSHT & 4.5 & 5.2 & 6.6 & 7.4 & 23.8  \\  \hline 
\centering Uncorrelated case \\ (CMS 7~TeV jets) & MSHT & 4.0 & 6.4 & 7.2 & 6.4 & 24.0  \\ \hline 
\centering Uncorrelated case \\ (ATLAS 7~TeV jets) & MSHT & 4.6 & 5.5 & 5.2 & 6.4 & 21.6  \\  \hline 

  \end{tabular}
\caption[]{$\chi^2$ for the ATLAS 8~TeV $t\bar{t}$ lepton + jets distributions when added to the reduced fit, before fitting with PDF4LHC15 PDFs read in, after fitting but including all 4 distributions in an ``uncorrelated'' manner and after fitting maintaining all statistical and systematic correlations between the distributions. The effect of double weighting small datasets and of the inclusion of different jet datasets in the reduced fit is also shown, with the LHC jet dataset included in brackets.}
\label{tab:ttbar_chisqs}                                
\end{table}

There are two potential explanations for these differences. The first is a methodological difference between the PDF fitting groups, NNPDF being based on a neural network, must divide each dataset into training and validation to prevent over-fitting. However, such a split is infeasible for small datasets and so all data are placed in the training, this potentially up-weights smaller datasets, such as the ATLAS $t\bar{t}$ rapidity distributions. To investigate this further a MSHT reduced fit was performed in which small datasets were double-weighted to attempt to approximately mimic this. This reveals a notable improvement in the fit to the ATLAS $t\bar{t}$ $y_t$ and $y_{tt}$ distributions, seen in the fourth row of Table~\ref{tab:ttbar_chisqs}, matching the qualitative pattern seen in NNPDF. In addition, beyond the effect on the ATLAS $t\bar{t}$ rapidity distributions, this may also be relevant for some of the differences in fit quality seen in the benchmarking of the reduced fits in Table~\ref{tab:reducedfit_chisqs_fit}, where it was remarked that the NNPDF reduced fit was able to fit smaller datasets better than the MSHT or CT reduced fits (with a slight trade-off for the fit quality of larger data sets). An example of such a dataset is the CMS 7~TeV electron asymmetry data, when double-weighted its fit quality improves markedly in the MSHT reduced fit - from $\chi^2/N_\mathit{pts} \sim 1.3$ to 0.9, more consistent with the $\sim~0.75$ in the NNPDF reduced fit and so potentially explaining this difference in the reduced fit comparison.

%\begin{table}[t]
%\fontsize{8}{10}\selectfont 
%  \renewcommand\arraystretch{0.9}
%  \centering
%  \begin{tabular}{|c|c|c|c|>{\centering\arraybackslash}m{2.8cm}|}
%%  \renewcommand\arraystretch{0.9}
%  %\cline{1-1}
%  %\multicolumn{1}{|l|}{}  
%   \hline
%	\multicolumn{2}{|c|}{  Dataset($N_\mathit{pt}$)} & MSHT uncorrelated & NNPDF uncorrelated & MSHT uncorrelated double weight \\ \hline 
%	\multicolumn{2}{|c|}{ Total} & 2314.1 & 2731.4 & 2313.3  \\ \hline 
%	\multicolumn{2}{|c|}{ $\chi^2/N_\mathit{pt}$} & 1.15 & 1.20 & 1.15  \\ \hline
%	\multicolumn{2}{|c|}{ E866 $\sigma_{pd}/(2\sigma_{pp})$ (15)} & 9.5 & 5.2 & 9.2 \\ \hline 
%	\multicolumn{2}{|c|}{ CMS 7~TeV electron $A_{ch}$ (11)} & 14.2 & 8.2 & 10.2 \\ \hline 
%	\multirow{5}{2.5cm}{\centering ATLAS 8~TeV $t\bar{t}$ lepton + jet} & $p_t^T$ (8) & 3.8 & 7.2 & 4.2 \\ 
%	 & $y_t$ (5) & 8.4 & 4.3 & 5.8 \\ 
%	 & $y_{tt}$ (5) & 12.5 & 5.7 & 7.4 \\ 
%	 & $m_{tt}$ (7) & 6.4 & 2.4 & 6.5 \\ \cline{2-5}
%	 & total (25) & 31.2 & 19.6 & 23.9 \\ \hline 
%  \end{tabular}
%\caption[]{MSHT and NNPDF reduced fit $\chi^2$ including the ATLAS 8~TeV $t\bar{t}$ lepton + jets dataset added with all 4 distributions included in the ``uncorrelated'' manner. The final column shows the effect of double weighting the smaller datasets present - the E866 $\sigma_{pd}/(2\sigma_{pp})$, CMS 7~TeV electron $A_{ch}$ and ATLAS 8~TeV $t\bar{t}$ lepton + jet datasets. Note only relevant datasets included in the reduced fits are shown here for simplicity.}
%\label{tab:smalldatasets_doubleweighting}                                
%\end{table}

Nonetheless, for such differences to potentially play a role it suggests they may be changing the balance of pulls between different datasets on the PDFs, therefore possible tensions between datasets in this high-$x$ gluon region are a further explanation. In order to investigate this, reduced fits were run in which the LHC jet data included was varied from the default CMS 8~TeV jets, the effects of this upon the MSHT reduced fit are also shown in Table~\ref{tab:ttbar_chisqs}. It is clear that there is tension between the CMS 8~TeV jet data and ATLAS 8~TeV multi-differential $t\bar{t}$ data, once the former is removed the $y_t$, $y_{tt}$ fit qualities improve significantly to the extent they are now able to be fit well. This is perhaps unsurprising given it is known that the CMS 8~TeV jet dataset tends to pull the gluon up at high $x$ \cite{MSHT20,CT18,AbdulKhalek:2020jut}, whilst the $t\bar{t}$ rapidity distributions favour a lower gluon in this region \cite{Kadir:2020yml}. It is therefore broadly consistent with the pulls observed in the full global fits. Similar effects are also observed upon replacement of the CMS 8~TeV jets data in the default reduced fits with the CMS 7~TeV jets or ATLAS 7~TeV jets data, which both favour a reduced gluon in the region of interest for the $t\bar{t}$. As a result these are more consistent with the ATLAS 8~TeV lepton+jet $t\bar{t}$ rapidity distributions, which are then well fit. This demonstrates the effects of tensions in the high-$x$ gluon region on the fit quality of the ATLAS 8~TeV $t\bar{t}$ lepton + jets data. Moreover such tensions are also present within the jet datasets themselves, with the CMS 8~TeV jets $\chi^2$ worsening upon inclusion of the CMS or ATLAS 7~TeV jets data and vice versa (not shown). This provides a potential answer to long-standing questions about the different behaviour seen for this $t\bar{t}$ data by the different global fitting groups, with each group investigating this data originally in the context of different baseline fits, and so with different jet datasets and weights of data involved.

\section{Reduced fit Luminosity Comparison}\label{sec:reducedfit_lumis}

\begin{figure}[h]
\centering
\includegraphics[width=0.495\textwidth,trim= 0.2cm 0.2cm 0.1cm 0.2cm,clip]{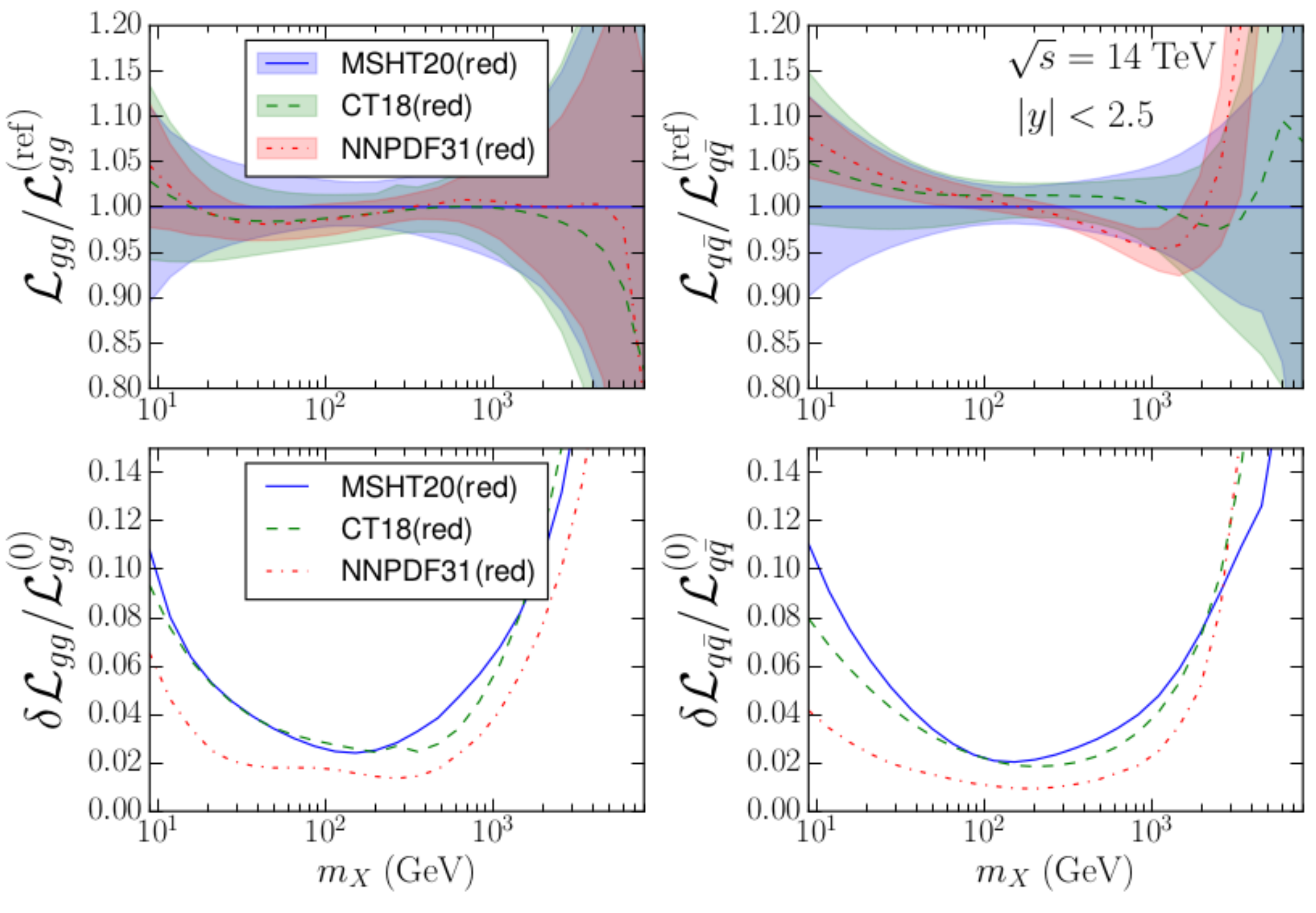}
\includegraphics[width=0.495\textwidth,trim= 0.2cm 0.2cm 0.1cm 0.2cm,clip]{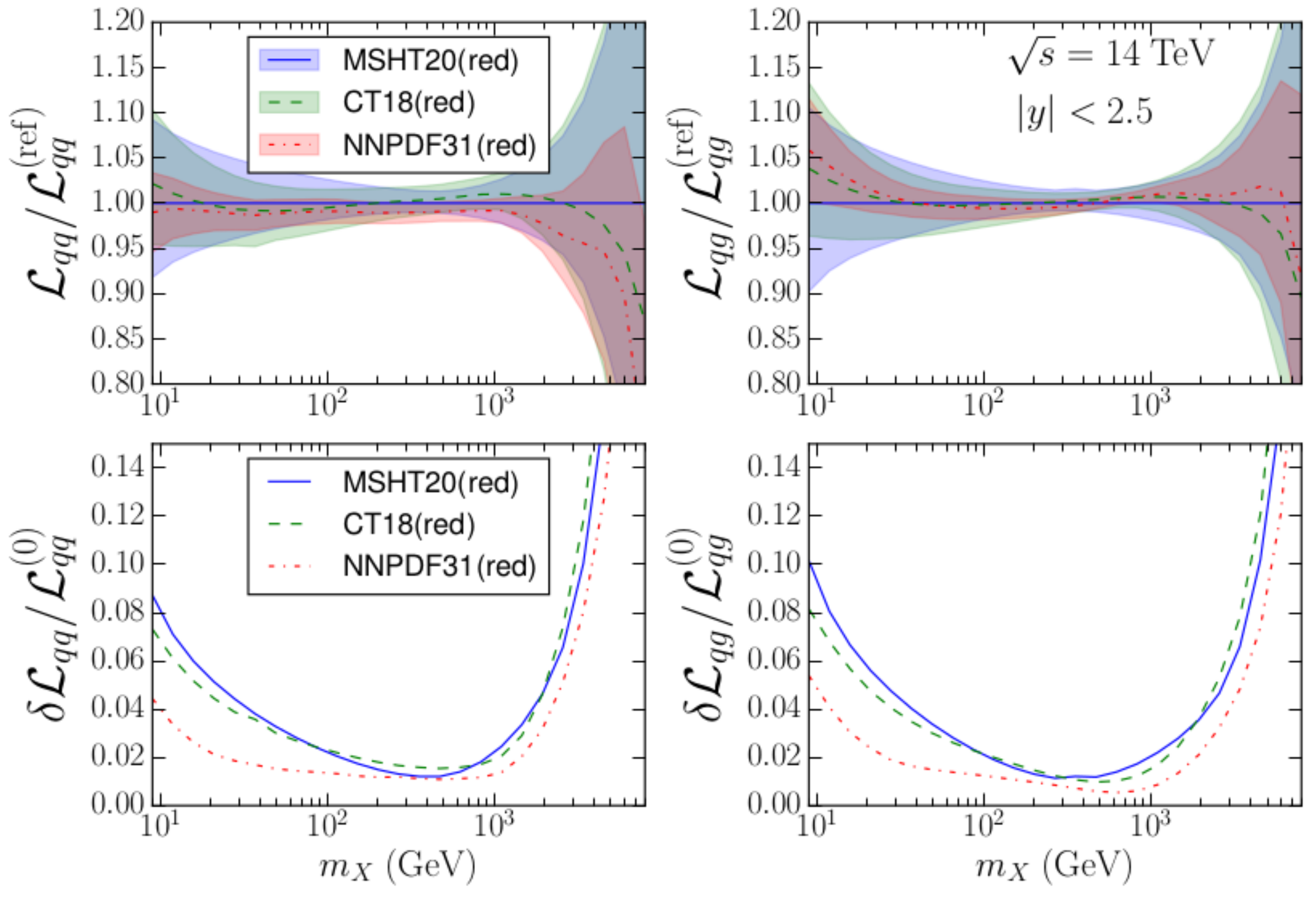}
\caption{Comparison of the luminosities of the 3 reduced fits for gluon-gluon, quark-antiquark, quark-quark and quark-gluon (going from left to right). The absolute luminosities relative to that of the MSHT reduced fit are shown in the upper plots, whilst the uncertainties in the luminosities are shown in the lower plots. A rapidity cut of $\lvert y \rvert < 2.5$ is applied for the luminosities.}
\label{fig:Reducedfit_lumi_comp}
\end{figure}

Overall it should be noted therefore that whilst some differences exist between the reduced fits, these have been reduced significantly in the process of the benchmarking. We have therefore seen that, provided the same data and theory settings are adopted (as far as possible), the resulting PDFs agree well within errors within the data regions, the benchmarking can therefore be considered a success. This is emphasised by comparing the reduced fit luminosities. Here we present a comparison of the gluon-gluon, quark-antiquark, quark-quark and quark-gluon luminosities for the three reduced fits, including both the ratios of their central values and their uncertainties, shown in Figure~\ref{fig:Reducedfit_lumi_comp}. There is excellent agreement between the reduced fits at the level of the parton-parton luminosities, with the important gluon-gluon luminosities agreeing within uncertainties across the entire $m_X$ range. Similar very good agreement is also seen in the quark-quark and quark-gluon luminosities. The agreement of the quark-antiquark luminosity is also good, however the NNPDF reduced fit luminosity is on the edge of the uncertainty bands for a small portion of the $m_X$ range, this perhaps signals some slight remaining difference in the quark flavour decomposition. Nonetheless, the overall agreement is very good, indicating the reduced fits are well understood and a good baseline for benchmarking. Moreover, the uncertainties in the luminosities are also similar between all three groups, particularly in the most constrained regions where data is available.

%\begin{figure}[h]
%\centering
%%\includegraphics[width=0.5\textwidth,trim= 0.5cm 0.5cm 0.3cm 0.4cm,clip]{figures/lumiplot-reduced_gg_qqb_Tom_new_July21_rapcut2p5.pdf}
%\includegraphics[width=0.5\textwidth,trim= 0.5cm 0.5cm 0.3cm 0.4cm,clip]{figures/lumiplot-reduced-July21-rap2p5cut-new3-ggqqb.pdf}
%\includegraphics[width=0.5\textwidth,trim= 0.5cm 0.5cm 0.3cm 0.4cm,clip]{figures/lumiplot-reduced_qq_qg_Tom_new_July21_rapcut2p5.pdf}
%\caption{Effect of the charm hadrons to muon branching ratio (BR) $B (D \rightarrow \mu)$ on the total strangeness of the MSHT reduced fit, relative to the MSHT20 global fit baseline.}
%\label{fig:Reducedfit_lumi_comp}
%\end{figure}

%\begin{figure}[h]
%\centering
%\includegraphics[width=0.6\textwidth,trim= 0.2cm 0.2cm 0.1cm 0.2cm,clip]{figures/lumiplot-reduced-July21-rap2p5cut-new6-ggqqb.pdf}
%\caption{Comparison of the luminosities of the 3 reduced fits for gluon-gluon (left) and quark-antiquark (right), the absolute luminosities relative to that of the MSHT reduced fit are shown in the upper plots, whilst the uncertainties in the luminosities are shown in the lower plots.}
%\label{lumis_reducedvsglobal_ggqqbar}
%\end{figure}
%\begin{figure}[h]
%\centering
%\includegraphics[width=0.6\textwidth,trim= 0.2cm 0.2cm 0.1cm 0.2cm,clip]{figures/lumiplot-reduced-July21-rap2p5cut-new6-qqqg.pdf}
%\caption{Comparison of the luminosities of the 3 reduced fits for quark-quark (left) and quark-gluon (right), the absolute luminosities relative to that of the MSHT reduced fit are shown in the upper plots, whilst the uncertainties in the luminosities are shown in the lower plots.}
%\label{lumis_reducedvsglobal_qqqg}
%\end{figure}

\section{Conclusion}\label{sec:conclusions}

In conclusion, we have demonstrated the benchmarking of the PDFs produced by the three main global fitting groups: CT, MSHT and NNPDF. This is a vital exercise required to understand the differences arising in the global fits and to provide a combination PDF set for use at the LHC in Run III and beyond: PDF4LHC21. This article has served to provide an update on this effort. Benchmarking has been achieved through the use of ``reduced fits'' to a small subset of datasets treated in (nearly) identical ways by all three groups and with common theory settings as far as possible. These reduced fits were compared to the global fits and then used to identify and understand any differences. This comparison has been carried out both at the level of the PDFs themselves, through their central values and uncertainties, and through the comparison of the $\chi^2$ fit qualities obtained for different datasets. Specific disagreements observed in the reduced fits were studied in more detail, in particular concerning differences in the strangeness of the reduced fit PDFs and in the high-$x$ gluon, the latter being of particular interest to global fits. The reduced fit environment has enabled such differences to be largely resolved. In particular the question of the fit quality to different distributions of the ATLAS 8~TeV $t\bar{t}$ lepton + jets data has been determined to be a reflection of tensions between datasets in the high-$x$ gluon region and the different weights assigned to them in the different global fit environments. As a result of the benchmarking, the reduced fit PDFs are now in very good agreement, as emphasised in the comparison of the PDF luminosities, and so the benchmarking can be considered a success.

The work presented in this article will continue within the PDF4LHC combination group, with the final aim of producing a benchmarking paper and a new PDF4LHC combination set, both key components of future phenomenological programs at the LHC.

%You must include a conclusion.
%
%
%Equations should be centered; multi-line equations should be aligned.
%\begin{equation}
%H = \sum_{j=1}^N \left[J (S^x_j S^x_{j+1} + S^y_j S^y_{j+1} + \Delta S^z_j S^z_{j+1}) - h S^z_j \right].
%\end{equation}

\section*{Acknowledgements}
T.C. would like to acknowledge the work towards this article by the whole of the PDF4LHC21 combination group and wishes to thank all of those involved in this work and many useful discussions, including in the wider PDF4LHC working group on PDFs and related issues. In particular, special thanks are given to A. Cooper-Sarkar, L. Harland-Lang, T. Hobbs, T.-J. Hou, J. Huston, P. Nadolsky, E. Nocera, J. Rojo and R. Thorne for providing tables, plots, fits, comments and other contributions used in this work.

%Acknowledgements should follow immediately after the conclusion.

%In the list of references, cited papers \cite{1931_Bethe_ZP_71} should include authors, title, journal reference (journal name, volume number (in bold), start page) and most importantly a DOI link. For a preprint \cite{arXiv:1108.2700}, please include authors, title (please ensure proper capitalization) and arXiv link. If you use BiBTeX with our style file, the right format will be automatically implemented.
%
%All equations and references should be hyperlinked to ensure ease of navigation. This also holds for [sub]sections: readers should be able to easily jump to Section \ref{sec:another}.

% TODO: include author contributions
%\paragraph{Author contributions}
%This is optional. If desired, contributions should be succinctly described in a single short paragraph, using author initials.

% TODO: include funding information
\paragraph{Funding information}
T. C. would like to thank the Science and Technology Facilities Council (STFC) for support via grant awards ST/P000274/1 and ST/T000856/1. Other contributors to this article would like to acknowledge support from STFC via grant award ST/L000377/1, from the Leverhulme Trust, the U.S. Department of Energy under  Grant No. DE-SC0010129, the U.S. National Science Foundation under Grant No. PHY-2013791, the U.S. Department of Energy under Grant No.~DE-AC02-07CH11359 and the U.S. National Science Foundation under Grant No. 2012165.

%T. C. and R. S. T. would like to thank the Science and Technology Facilities Council (STFC) for support via grant awards ST/P000274/1 and ST/T000856/1. L. H. L. thanks STFC for support via grant award ST/L000377/1. P. N. ... The work on the presented results is supported in part by the US Department of Energy under  Grant No. DE-SC0010129 (at SMU), by the U.S. National Science Foundation under Grant
%No. PHY-2013791 (at MSU)

%Authors are required to provide funding information, including relevant agencies and grant numbers with linked author's initials. Correctly-provided data will be linked to funders listed in the \href{https://www.crossref.org/services/funder-registry/}{\sf Fundref registry}.

\begin{appendix}

\bibliography{references.bib}
%\end{verbatim}
%at the end of your document. If you are not using our \LaTeX template, please still use our bibstyle as
%\begin{verbatim}
%\bibliographystyle{SciPost_bibstyle}
%\end{verbatim}
%in order to simplify the production of your paper.

% TODO:
% Provide your bibliography here. You have two options:

% FIRST OPTION - write your entries here directly, following the example below, including Author(s), Title, Journal Ref. with year in parentheses at the end, followed by the DOI number.
%\begin{thebibliography}{99}
%\bibitem{1931_Bethe_ZP_71} H. A. Bethe, {\it Zur Theorie der Metalle. i. Eigenwerte und Eigenfunktionen der linearen Atomkette}, Zeit. f{\"u}r Phys. {\bf 71}, 205 (1931), \doi{10.1007\%2FBF01341708}.
%\bibitem{arXiv:1108.2700} P. Ginsparg, {\it It was twenty years ago today... }, \url{http://arxiv.org/abs/1108.2700}.
%\end{thebibliography}

% \bibliographystyle{SciPost_bibstyle} % Include this style file here only if you are not using our template
%\bibliography{SciPost_Example_BiBTeX_File.bib}

\end{appendix}

\nolinenumbers

\end{document}